\documentclass[aps,twocolumn,superscriptaddress,showpacs]{revtex4}
\usepackage{epsf,latexsym}
\usepackage{times}

\newcommand{\bra}[1]{\langle #1 |}
\newcommand{\ket}[1]{| #1 \rangle}

\newcommand\R{{\mathrm {I\!R}}}

\newcommand\h{{\cal H}}

\newcommand{\be}{\begin{equation}}
\newcommand{\ee}{\end{equation}}

\def\CC{{\rm\kern.24em \vrule width.04em height1.46ex depth-.07ex
    \kern-.30em C}}
\def\P{{\rm I\kern-.25em P}}

\def\RR{{\rm
         \vrule width.04em height1.58ex depth-.0ex
         \kern-.04em R}}

\def\bbbc{{\mathchoice {\setbox0=\hbox{$\displaystyle\rm C$}\hbox{\hbox
to0pt{\kern0.4\wd0\vrule height0.9\ht0\hss}\box0}}
{\setbox0=\hbox{$\textstyle\rm C$}\hbox{\hbox
to0pt{\kern0.4\wd0\vrule height0.9\ht0\hss}\box0}}
{\setbox0=\hbox{$\scriptstyle\rm C$}\hbox{\hbox
to0pt{\kern0.4\wd0\vrule height0.9\ht0\hss}\box0}}
{\setbox0=\hbox{$\scriptscriptstyle\rm C$}\hbox{\hbox
to0pt{\kern0.4\wd0\vrule height0.9\ht0\hss}\box0}}}}

\def\bbbz{{\mathchoice {\hbox{$\sf\textstyle Z\kern-0.4em Z$}}
{\hbox{$\sf\textstyle Z\kern-0.4em Z$}}
{\hbox{$\sf\scriptstyle Z\kern-0.3em Z$}}
{\hbox{$\sf\scriptscriptstyle Z\kern-0.2em Z$}}}}

\newcommand{\putfig}[2]{$$\leavevmode\hbox{\epsfxsize=#2 cm
   \epsffile{#1.eps}}$$}

\begin{document}
\title{Quantum entangling power of adiabatically connected hamiltonians }
\author{Alioscia Hamma$^{1,3,4}$, Paolo Zanardi$^{1,2}$}
\address{ Institute for Scientific Interchange (ISI), Villa
Gualino, Viale Settimio Severo 65, I-10133 Torino, Italy
\\
$^2$ Department of Mechanical Engineering, Massachusetts
Institute of Technology, Cambridge Massachusetts 02139
\\ 
$^3$ Dipartimento di Scienze Fisiche, Universit\`a
Federico II di Napoli, Via Cintia ed. G,\\
$^4$Istituto
Nazionale di Fisica Nucleare (INFN), sezione di Napoli
}
\begin{abstract}
The space of quantum Hamiltonians has a natural partition in
classes of operators that can be adiabatically deformed into each
other. We consider parametric families of Hamiltonians acting on a
bi-partite quantum state-space. When the different Hamiltonians in
the family fall in the same adiabatic class one can manipulate
entanglement by  moving through energy eigenstates corresponding
to different value of the control parameters. We introduce an
associated notion of adiabatic entangling power. This novel
measure is analyzed for general $d\times d$ quantum systems and
specific two-qubits examples are studied
\end{abstract}
\maketitle
\section{ Introduction}
Adiabatic evolutions represent a very special class of quantum evolutions, nevertheless
they allow for a broad set of quantum state manipulations. In particular
a big deal of activity has been recently devoted to the study of adiabatic techniques
for Quantum Information Processing \cite{QIP}.

The notion of adiabatic quantum computing  emerged as an novel
intriguing paradigm for the development of  efficient quantum
algorithms \cite{aqc},\cite{berk},\cite{berk1}. In this approach
information e.g., the solution of an hard combinatorial problem,
is encoded in  the ground state of a properly designed many-qubits
Hamiltonian $H_f.$ This ground-state is  then generated by letting
the system evolve in adiabatic fashion from the ground state of a
simple initial Hamiltonian $H_0$ \cite{aqc}. In view of the
adiabatic theorem ( see e.g.,  \cite{mess}) the crucial property
which governs the scaling behaviour of the computational time is
the spectral gap i.e., energy difference between the ground and
the first excited state. The larger the gap the faster the
computation can be.

In adiabatic quantum computing  as defined in Ref. \cite{aqc} the parametric family of Hamiltonians has the simple form of a convex combination of
$H_0$ and $H_f;$ one can also consider more general  family of Hamiltonians and more complex paths in the control parameter space.
For example in the so-called {\em geometric quantum computation} \cite{GQC}
one considers {\em loops} in the control space of a non-degenerate set of Hamiltonians
to the purpose of controlled   Berry phases generation \cite{berry}.
When even the non-degeneracy constraint is lifted one and high-dimensional eigenspaces
are allowed, one is led to consider non-abelian
holonomies which  mix non-trivially the ground-states of the system. This latter method,
which  provides a general approach to QIP as well, is termed {\em holonomic quantum computation}
\cite{HQC}

In this paper we shall investigate how one can adiabatically
generate quantum entanglement \cite{unan},\cite{pz}. The idea is a
simple one. One first prepares a bi-partite  quantum system in one
of its eigenstates e.g., the ground-state, and then drives the
control parameters of the system Hamiltonian along some path. If
this path is  adiabatic the system will stand at any any time in
the corresponding eigenstate. In general eigenstates associated to
different control parameters will have different entanglement,
therefore the described dynamical process will result in a
protocol for entanglement manipulation. We would like to
characterize a parametric family of Hamiltonians in terms of its
capability of entanglement generation according the above
protocol. In this paper we  will  focus on bi-partite e.g.,
two-qubits, quantum systems. The aim will be, given an Hamiltonian
family, to characterize its entangling capabilities by means of
adiabatic manipulations.

\section{Adiabatic connectibility}

Let us start by a few simple general considerations about adiabatically connectible Hamiltonians.
We would like to understand how the space of Hamiltonians over ${\cal H}\cong\CC^D,$
splits in classes of  elements that can be adiabatically deformed into each other.

{\em Definition} Two Hamiltonians $H_0$ and $H_1$ are adiabatically connectible
if its exists a continuous family of Hamiltonians $\{H_t\}_{t\in[0,1]}
$ such that i) $H(0)=H_0$ and $H(1)=H_1,$ ii)
the degeneracies of the spectra of the $H_t'$s do not depend on $t.$

The notion of adiabatic deformability of Hamiltonians is an important concept
in many-body and field theory quantum systems. Indeed when  two Hamiltonians
can be connected in this way they share several properties e.g., ground-state degeneracy,
quasi-particle quantum numbers,$\ldots,$
so that  in many respects they can be regarded  as belonging to the same kind of universality class
\cite{PWA}. On the other an obstruction to such a process will be typically
associated to a some sort of quantum phase transition. Unconnectible Hamiltonians
show  qualitative different features. Since we will study how entanglement changes
while remaining in the same adiabatic class our  analysis can be regarded
as complimentary to the one of entanglement behaviour in quantum phase transitions
\cite{QPT}.

In the simple finite-dimensional case we are interested in one can prove the following

{\em{ Proposition 1}}.-- Two Hamiltonians $H_0$ and $H_1$ over
${\cal H}\cong\CC^D,$ are adiabatically connectible if and only if
they belong to the same connected component of the set of iso-degenerate Hamiltonians. 

{\em Proof.} 
Let $H_\alpha=\sum_{i=1}^R \epsilon_\alpha^i
\Pi_\alpha^i \,(\alpha=0,1)$ the spectral resolution of   $H_0$ and $H_1.$
We now order their eigenvalues in ascending order i.e., $\epsilon_\alpha^1<...<\epsilon_\alpha^R.$ We define two { vectors} $D_\alpha\,(\alpha=0,1)$
in $\R^R$ as follows $D_\alpha:=(\rm{tr} \Pi_\alpha^1,\ldots,\Pi_\alpha^R),$ where the components are ordered according to the corresponding eigenvalues.
The Hamiltonians  $H_0$ and $H_1$ belong  to the same connected component of the set of iso-degenerate hamiltonians
iff  $D_0=D_1.$ Iso-degeneracy is given by the weaker condition that it exists a permutation $P$ of $R$ objects such
that $(D_1)_i=(D_0)_{P(1)},\,(i=1,\ldots,R).$
It is an elementary fact that, given the two systems of
ortho-projectors $\{\Pi_\alpha^i\}_{i=1}^R, \,(\alpha=0,1)$ such
that Tr$\Pi_1^i=\mbox{Tr} \Pi_2^i,\,(i=1,\ldots,R),$ it exist a
(non-unique ) unitary $W$ such that $W
\,\Pi_0^i\,W^\dagger=\Pi_1^i,\,(i=1,\ldots,R).$ 
Let us introduce $R$ real-valued functions $\epsilon^i\colon [0,\,1]\mapsto \R$ 
 such that $\epsilon^i(0)=\epsilon_0^i,$ and $\epsilon^i(1)=\epsilon_1^i\, (i=1,\ldots,R).$ 
 In view of the ordering assumption  we can choose them to satisfy the no-crossing constraints $\epsilon^{i+1}(t)>\epsilon^{i}(t)\,
(i=1,\ldots,R-1).$
Consider now the following family of Hamiltonians $H(t)=\sum_{i=1}^R \epsilon^i(t) U_t \Pi_0^i U_t^\dagger,$
where the continuous unitary family $\{U_t\}_{t=0}^1$ is such that $U_0=\openone$ and $U_1=U.$
Clearly $H(0)=H_0$ and $H(1)=H_1.$ Moreover, for the very way they have been constructed, all the $H(t)$
belong to the same connected component of the set of iso-degenerate Hamiltonians of $H_0$ and $H_1.$
 This shows that the latter condition is sufficient in order that $H_0$ and $H_1.$ are adiabatically connectible.

Iso-degeneracy of $H_0$ and $H_1$ is also an obvious necessary condition for adiabatic  connectibility
because otherwise level crossing would necessarily occur. But level crossing would necessarily occur even if $D_0\neq D_1$ because, for some $t\in[0,\,1],$
and $1\le i\le R,$ it would be $\epsilon^{i+1}=\epsilon^{i}.$ This proves the necessity part of the Proposition. 

$\hfill\Box$

The role of the functions $\epsilon^i(t)$ in the Proof above is to map the spectrum of $H_0$ onto the one of $H_1$
whereas all the information about the eigenvectors is contained in the family of unitaries $U_t.$
By setting all the connecting functions $\epsilon^i_t/\epsilon_0^i$ to one, one gets a final Hamiltonian $\tilde{H}_1$
{\em iso-spectral} to $H_ 0$ having the same eigenvectors of $H_1.$
This latter remark is important for the following in that
it allows one to restrict to iso-spectral Hamiltonian families.
The actual spectrum structure e.g., the energy gaps, just imposes
an upper bound  over the speed at which the adiabatic deformation process can be  carried on.
Moreover in order to have a  one-to-one  correspondence between eigenvalues and eigenstates we shall
assume that our Hamiltonians are   {\em non-degenerate} i.e., $d_i=1,\,(i=1,\ldots,R).$
Notice that in Hamiltonian space the condition of non-degeneracy is a {\em generic} one.

The simplest case one can consider  is of course provided by two-level Hamiltonians
with eigenvalues $\epsilon_1$ and $\epsilon_2.$
Using the standard pauli matrices, one can write $H =\epsilon_S \openone +\epsilon_A \vec{n}\cdot\vec{\sigma}\,
(\epsilon_S:= (\epsilon_1+ \epsilon_2)/2,\,\epsilon_A:=( \epsilon_1- \epsilon_2)/2$
Here we have just two possibility 1) $\epsilon_A=0$ the
Hamiltonian is a rescaled identity and we have just one degree of
freedom ii) $\epsilon_A\neq 0;$ all possible operators of this
kind are then parameterized by a triple $(\epsilon_S, \epsilon_A,
\vec{n})$ where $\epsilon_A\in \RR,\, \epsilon_A\in\RR-\{0\}$ and
$\vec{n}\in S^2\cong SU(2)/U(1).$
For each of the two iso-degeneracy classes above there is just one connected component
i.e., any  the non (totally) degenerate Hamiltonian is adiabatically connectible 
any other  non (totally) degenerate Hamiltonian. Notice that this latter statement holds
for any dimension of ${\cal H}.$
%

\section{ Adiabatic entangling power.}

We move now to introduce our definition of adiabatic entangling power. Let ${\cal H}\cong
\CC^d\otimes\CC^d$ be a bi-partite quantum state space. We
consider a  family $\cal F$ of non-degenerate Hamiltonians
over $\cal H,$
${\cal F}_H:=\{H(\lambda)\,/\, \lambda \in{\cal M}\}
$
where $\cal M$ is a $n$-dimensional compact and connected manifold.
The points of $\cal M$ are to be seen as dynamically controllable parameters.
Let $E\colon {\cal H}\rightarrow \RR^+_0$ be a measure of bi-partite pure state
entanglement over $E$ e.g., von Neumann entropy of the reduced density matrix.
If $H(\lambda)=\sum_{i=1}^{d^2} \varepsilon_i |\Psi_i(\lambda)\rangle\langle \Psi_i(\lambda)|$
is the spectral resolution of an element of $\cal F$ we define
the {\em adiabatic entangling power} of $\cal F$ by
\begin{equation}
e({\cal F}_H):= \max_{i} \,\sup_{\lambda,\lambda^\prime}
|E(|\Psi_i(\lambda)\rangle)-E(|\Psi_i(\lambda^\prime)\rangle)|
\label{ep}
\end{equation}
$(i=1,\ldots,d^2,\,\lambda,\lambda^\prime\in{\cal M})$

We will assume  that it exists $H_{\lambda_0}\in{\cal F}_H$ such that the
associated eigenvectors are all{ \em product states}.
Let us stress once again that the physical idea underneath these definitions is quite simple:
one starts from  the (unentangled) eigenvectors of $H_{\lambda_0}$
then by adiabatically driving the control parameters $\lambda$
the states $|\Psi_i(\lambda)\rangle$ can be reached.
If $\lambda^*$ denotes the point of where the maximum (\ref{ep})
is achieved ($\cal M$ is compact) any adiabatic path connecting $\lambda_0$
to $\lambda^*$ realizes an {\em optimal} entanglement generation procedure
within the family ${\cal F}_H.$

An explicit evaluation of (\ref{ep}) is, for a general $\cal F,$
quite a difficult task. In the light of the observations after
Proposition 1,  we can, without loss of generality,   consider
only the case in which $\cal F$ is an {\em iso-spectral} family of
{non-degenerate } Hamiltonians. Let ${\cal F}_U\subset {\cal
U}(\CC^d\otimes\CC^d)$ be a set (compact and connected) of unitary transformations
containing the identity. The isospectral family is
\begin{equation}
{\cal F}_H:=\{ U\,H_0 U^\dagger\,/\,
\,U\in{\cal F}_U\}
\label{family}
\end{equation}
where $H_0=\sum_{i=1}^{d^2}
\varepsilon_i|\Psi_i\rangle\langle\Psi_ i|,\; i\neq j\Rightarrow
\varepsilon_i\neq\varepsilon_j,$ and  the $|\Psi_i\rangle$'s are
an orthonormal basis of product states. Moreover one can also
restrict herself  to {\em ground-state} entanglement i.e., to
consider the entanglement contents of just the eigenvector
$|\Psi_0\rangle$ corresponding to the minimum energy eigenvalue.
If this is the case one can forget about the maximization over the
eigenvalue index $i$ in Eq. (\ref{ep}). The ground state of
$H(\lambda)$ ($H_0$) will be denoted as $|\Psi_0(\lambda)\rangle$
($|\Psi_0\rangle$). For an iso-spectral family as in Eq.
(\ref{family}) we will use the notation $e({\cal F}_U).$

The adiabatic entangling power (\ref{ep}) induces, for the  class of Hamiltonian families (\ref{family}) the following 
real-valued function over the subsets ${\cal F}_U$ of ${\cal U}(\CC^d\otimes\CC^d).$
\begin{equation}
e({\cal F}_U)=  \max_i \sup_{ U\in{\cal F}_U}
E[U |\Psi_i\rangle].
\label{ep1}
\end{equation}
It is important to stress that this expression has the physical meaning of entanglement achievable  
by adiabatically manipulating the parametersi, living in a manifold ,say, $\cal M$, on which the $U$'s in ${\cal F}_U$ depend.
Indeed, for an iso-spectral Hamiltonian family (\ref{family}) the adiabatic evolution operator corresponding
to the path $\gamma\colon [0,\,T]\mapsto \cal M$ is given by the product of three different kinds of contributions
$U_{ad} (\gamma)=U(\gamma(T))\, e^{-i H_0 T}\,U_B(\gamma).
$
The first term $U(\gamma(T))$ is simply the unitary corresponding
to the end-point of the path $\gamma.$ Due to the adiabatic
theorem an initial eigenstate $|\Psi_i\rangle$ is indeed mapped,
up to a phase, onto the final eigenstate
$U(\gamma(T))|\Psi_i\rangle.$ The  second factor in $U_{ad}$
is clearly just the dynamical phase associated with $H_0$ whereas
the third is an operator taking into account the geometric
contribution to the phase accumulated by the eigenvectors
$U_B(\gamma)= \sum_{i=1}^{d^2}e^{i\phi_g(\gamma)}|\Psi_i\rangle\langle \Psi_i|
$
in which $\phi_g(\gamma)=i\int_\gamma \langle\Psi_i(\lambda)|d|\Psi_i(\lambda)\rangle$
are the Berry's phases associated to $\gamma.$
Notice in passing that when $\gamma$ is a loop i.e., $\gamma(0)=\gamma(T)=\lambda_0$
then $U(\gamma(T))=\openone.$
As far as the adiabatic entangling power (\ref{ep}) is concerned
the phases can be obviously neglected. 

The adiabatic entangling power is invariant under left (and not
right in general) multiplication by bi-local unitary operators
i.e., $e({\cal F}_U)= e((U_1\otimes U_2){\cal F}_U),\,\forall
U_1,U_2\in{\cal U}(d).$ This implies that, as far adiabatic
entangling capabilities are concerned, a unitary family ${\cal
F}_U$ can be always considered closed under the
left-multiplication by local unitary operators.

We want now to establish a connection between the adiabatic entangling power (\ref{ep1})
and a variation of entangling power $e_p^{(av)}$ of -bi-partite unitaries introduced
in Ref. \cite{dur}[For a different definition, based on {\em average} entanglement production,
see also \cite{ZZF}]. In this paper we define $e_p(U)$ as the {\em maximum} entanglement obtainable
by the action of $U$ over all possible product states i.e., 
$e_p(U)=\sup_{\psi_1, \psi_2} E[U|\psi_1\rangle\otimes|\psi_1\rangle].$ 

Since the $|\Psi_i\rangle'$s are  by hypothesis  product states one clearly has
$E[U |\Psi_i\rangle]\le \sup_{\psi_1,\psi_2} E[U |\psi_1\rangle\otimes|\psi_1\rangle],$
Therefore  one obtains  the upper bound
\begin{equation}
e({\cal F}_U)\le \sup_{U\in{\cal F}_U} e_p( U)
\label{bound}
\end{equation}
In some circumstances one can get the equality.

{\em Proposition 2}.-- Suppose  that the unitary family ${\cal
F}_U$ is such that for all $U_1, U_2\in{ U}({d})$ one has ${\cal
F}_U (U_1\otimes U_2)\subset {\cal F}_U$ i.e., the  family is
closed also under right multiplication of bi-local operators.  It
follows that that the adiabatic entangling power coincides with
the supremum over ${\cal F}_U$ of the entangling power $e_p(U).$

{\em Proof.}
It is straightforward
\begin{eqnarray}
e({\cal F})&=&\max_i \sup_{U\in{\cal F}_U, U_1, U_2}E[U  (U_1\otimes U_2)|\Psi_i\rangle] \nonumber \\
&=&\sup_{U\in{\cal F}_U, \psi_1, \psi_2}  E[U |\psi_1\rangle\otimes|\psi_2\rangle]
\ge \sup_{U\in{\cal F}_U}e_p(U).
\end{eqnarray}
Therefore using Eq. (\ref{bound}) one obtains
$e({\cal F})=\sup_{U\in{\cal F}_U} e_p(U_\lambda).$
Notice also that for such a family the maximization over the eigenvalue index $i$ in Eq. (\ref{ep}) is irrelevant.
$\hfill\Box$

\section{Examples}

We will now illustrate the use of the general notions introduced so far
by considering in a detailed fashion some concrete Hamiltonian
families acting on a two-qubits space. Before doing that let us
remind a few basic facts about two-qubits entanglement in pure
states. We  denote the standard  product basis by
$|\Psi_i\rangle,\,(i=1,\ldots,4) $ and consider a generic
two-qubits state $\ket{\Phi}=U \ket{\Psi}=\sum_{i=1}^{4} a_i
\ket{\Psi_i}$. The eigenvalues of the associated reduced density
matrix
 are given by $\lambda=(1+\sqrt{1-4C^2})/2$ and $1-\lambda$, where
$C^2=\left| a_1 a_2- a_3 a_4\right| ^{2}$ and $2\,C$ is the so called
'concurrence'. The entanglement measure is given by $E=-\left[
\lambda \log_2 \lambda +\left( 1-\lambda \right) \log_2 \left(
1-\lambda \right) \right] $. Since $\frac{dE}{d\lambda}<0$,
finding the maximum possible entanglement for the output state
$\ket{\Phi}$ means minimizing $\lambda$, or, which is the same,
maximizing $C^2$. The state $\ket{\Phi}$ is maximally entangled
for $\lambda=\frac{1}{2}$, or $C^2=\frac{1}{4}$.

{\em Example 0.}-- It is useful to start with an example of a
two-qubits  Hamiltonian family with {\em zero} adiabatic
entangling power. Let $H(\lambda)=\sum_{\alpha=x,y,z}
\lambda_\alpha \sigma_\alpha\otimes\sigma_\alpha,$ where the
$\lambda'$s are such that the corresponding hamiltonian is always
not degenerate One has that
$[H(\lambda),\,H(\lambda^\prime)]=0,\,(\forall\lambda,\lambda^\prime),$
then all the elements of the family can be simultaneously
diagonalized. The joint eigenvectors are clearly given by the
Bell's basis
$|\Phi^\pm\rangle:=1/\sqrt{2}(|00\rangle\pm|11\rangle),\,|\Psi^\pm\rangle:=1/\sqrt{2}(|10\rangle\pm|01\rangle).$
Entanglement in the eigenstates is therefore maximal an cannot be
changed by varying the control parameters $\lambda.$ Analogously
one can easily build examples of Hamiltonian families having joint
constant eigenvectors given by products.

\begin{figure}
\putfig{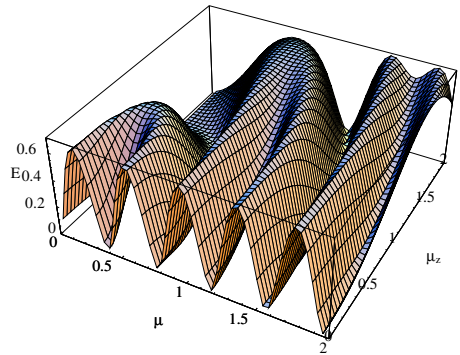}{6} \caption{(Color online) Entanglement generated by the
hamiltonian of the example 1 for the input state $\ket{01}$ as a
function of the parameters $\mu,\mu_z$.} \label{fig1}
\end{figure}

{\em Example 1.}--
The  non-degenerate Hamiltonian we  consider is the following
\begin{equation}
H_0= \lambda_1 \sigma_z\otimes \openone + \lambda_2  \openone
\otimes\sigma_z,\quad (\lambda_1\neq\lambda_2)
\end{equation}
The eigenvectors are given by the standard product basis.
We introduce  the family of unitaries $U(\mu,\mu_z)=\exp[i K(\mu,\mu_z)]$
where
\begin{equation}
K(\mu,\mu_z):=\mu \sigma^+\otimes\sigma^-+\bar{\mu} \sigma^-\otimes\sigma^+
+\mu_z(\sigma_z\otimes \openone - \openone \otimes\sigma_z)
\label{trans}
\end{equation}
and the associated iso-spectral family
$H(\mu,\mu_z):=U(\mu,\mu_z) H_0 U(\mu,\mu_z)^\dagger.$
The Hilbert space is given by $\h={\rm span}\left\{\ket{00}
,\ket{01},\ket{10} ,\ket{11} \right\}$
 and we can split it in the two subspaces $\h_0={\rm span}\left\{\ket{00}
,\ket{11}\right\}$
 and $\h_1={\rm span}\left\{ \ket{01} ,\ket{10}\right\}$, where obviously
$\h=\h_0\oplus \h_1$.

The evolution operator $U$ is the identity on $\h_0$ while it is a
straightforward exercise to verify that on $\h_1$ it yields:
$U\ket{01} \equiv \ket{\xi}\equiv a \ket{01} +b\ket{10}$ and $U\ket{10}
\equiv \ket{\zeta} \equiv-
\overline{b}\ket{01} +\overline{a}\ket{10}$, where $a=\cos \theta
+\frac{2i\sin \theta }{\theta }\mu _{z}$,
 $b=\frac{4i\sin \theta }{\theta }\overline{\mu}$ and
  $\vec \theta \equiv 2(\mu +\overline{\mu},i(\mu -\overline{\mu }),\mu _{z})$.
 For the
generic state $\ket{\Psi}= \alpha\ket{01} +\beta\ket{10}
+\gamma\ket{00}+ \delta\ket{11}$ one has 
$C^2=\left| xy-\gamma\delta\right|^{2},$ where $x=\alpha
a-\beta\overline{b}$ and $y=\alpha b+\beta\overline{a}$.

For  $\ket{01}$ the evolved state is $\left| \xi \right\rangle
=a\ket{01} +b\ket{10} $ and its reduced density
matrix is obviously $\rho =\rm{diag}(\left| a\right| ^{2},\left|b\right|^{2})$
whose eigenvalues are $\left|a\right|^{2}$ and $1-\left| a\right| ^{2}$.
The condition to obtain a maximally entangled state is hence $\left|a\right|^{2}=\frac{1}{2}$, that is,
$
\sin ^{2}\theta =\frac{1}{2}\left[ 1+\left( \frac{\mu _{z}}{2\left| \mu
\right| }\right) ^{2}\right]   \label{eq1}$. This equation admits (at
least) one solution iff
$\left|\mu _{z}\right| \leq 2\left| \mu\right| .$ Thus a maximally
entangled state can be reached starting from either  $\ket{01} ,\ket{10}.$
In Fig.  \ref{fig1} is showed  the reachable entanglement from the input
state $|01\rangle$  
as a function of the parameters $\mu ,\mu _{z}.$ We see how moving in the parameter space to 
 higher values of $\mu_z$ spoils the reachibility of a maximally entangled state.

\begin{figure}
\putfig{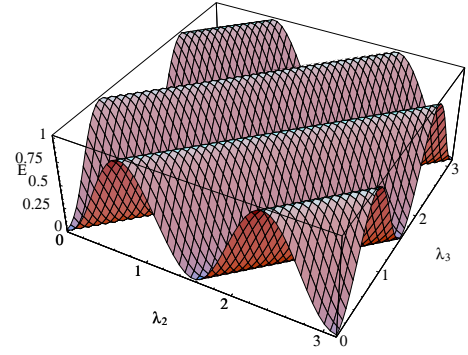}{6}
\caption{(Color online) Entanglement for the unitary $U=e^{i \sum_{i=1}^{3} \lambda_i
\sigma^i \otimes \sigma^i}$, with the input state
 $\frac{1}{\sqrt{2}} (\ket{\Psi_1} +i\ket{\Psi_2})$ as a  function of
$\lambda_2,\lambda_3$, with $\lambda_1=1$.}
\label{fig2}
\end{figure}

{\em Example 2}.--
Let us examine now the following unitary family:
$U=\exp(i \sum_{j=1}^{3} \lambda_j \sigma_j \otimes \sigma_j)
$
In the so-called magic basis $[
\ket{\Psi_1}=({\ket{00}+\ket{11}})/\sqrt{2},\ket{\Psi_2}=-i(\ket{00}-\ket{11})/\sqrt{2}
,\ket{\Psi_3}=
(\ket{01}-\ket{10})/\sqrt{2},\ket{\Psi_4}=-i(\ket{01}+\ket{10})/\sqrt{2}
]$
 (as well in the Bell basis) these unitaries are diagonal and read
$U=\sum_{k=1}^{4}e^{i h_k} \ket{\Psi_k} \bra{\Psi_k}$
 where $\{h_1=\lambda_1 -\lambda_2 +\lambda_3, h_2=\lambda_1 +\lambda_2
-\lambda_3, h_3=-\lambda_1 +\lambda_2 +\lambda_3,
 h_4=-\lambda_1 -\lambda_2 -\lambda_3\}.$ So in this basis the input state
is $\ket{\Psi}=\sum_{k}w_k\ket{\Psi_k}$ and the
  output state is $\ket{\Phi}=\sum_{k}w_k e^{-i h_k}\ket{\Psi_k}.$  The concurrence is given
   by $C^2=\sum_{k,l} (w_k e^{-i h_k})^2 (w^*_l e^{i h_l})^2$. Following
Ref \cite{kc}, we find that the
    maximum reachable concurrence is $C=\max_{k,l} |\sin(h_k-h_l)|$ and the
product input state which gives the
    best entangling capability as a  function of the parameters $\lambda_k$
is then $\frac{1}{\sqrt{2}} (\ket{\Psi_k} +i\ket{\Psi_l}).$
     So for instance a maximally entangled state can be reached from the
input state $\frac{1}{\sqrt{2}} (\ket{\Psi_1} +i\ket{\Psi_2})=|00\rangle$ for
      parameters such that $\lambda_3 - \lambda_2 = \pi/4$ (see Fig. \ref{fig2}).

Before passing to the conclusions we would like to show that the
first two-qubit Hamiltonian family associated with the unitaries
(\ref{trans}) can be used to generate a non trivial entangling
gate in an adiabatic fashion. 

{\em{Proposition 3}}
An adiabatic  loop in the parameter space $(\mu,\mu_z)\,(|\mu|^2+\mu_z^2=\mbox{const})$
gives rise to the diagonal  unitary mapping $|\alpha\beta\rangle\rightarrow
\exp(i\phi_{\alpha\beta})|\alpha\beta\rangle$
where, if $\gamma$ denotes the geometric contribution , $E_{\alpha\beta}$ the eigenvelues
and $T$ is the operation time, one has
$\phi_{01}= E_{01} \,T +\gamma,\,\phi_{10}= E_{01} \,T -\gamma,\,\phi_{00}= E_{00}\,T,\,
\phi_{11}= E_{11}\,T$
For $\phi_{01}+\phi_{10}-(\phi_{00}+\phi_{11})= -4\,T \neq 0 \mbox{mod}\, 2\pi$
the obtained transformation is equivalent to a controlled-phase-shift.

{\em{Proof.}}
Indeed it is easy to check that
i) by the adiabatic theorem the evolution has to be diagonal in the
product basis
ii) the geometric contribution of the states $|\alpha\alpha\rangle\,(\alpha=0,1)$
is zero ($\Leftarrow U(\mu,\mu_z)|\alpha\alpha\rangle=|\alpha\alpha\rangle$)
iii) In the one-qubit subspace spanned by $|{\bf{0}}\rangle:= |01\rangle$ and $|{\bf{1}}\rangle:=|10\rangle$
the  unitaries $e^{iK}$ with the $K$ defined in (\ref{trans}) look like
$U(\mu,\mu_z)=exp  [i(\mu \tilde{\sigma}^+ +\bar{\mu}\tilde{\sigma}^- +\mu_z\tilde{\sigma}_z)].$
This latter equation can be of course written as ${\bf{B}}\cdot \dot{\bf{\sigma}}$
where a fictious magnetic field ${\bf{B}}$ has been introduced.
One can then use the standard Berry-phase argument for a spin $1/2$ particle in
an adiabatically  changing magnetic field to claim that under a ${\bf{B}}$ going along an adiabatic loop,
 one has $|{\bf{0}}\rangle\mapsto e^{i\gamma} |{\bf{0}}\rangle$
and $|{\bf{1}}\rangle\mapsto e^{-i\gamma} |{\bf{1}}\rangle$. Here $\gamma$ denotes the standard geometric phase
i.e., proportional to the solid angle swept by ${\bf{B}}.$
The final equivalence claim stems from a known result in literature \cite{calarco}.
$\hfill\Box$

Of course the general fact that entangling gates  can be obtained  via adiabatic  manipulations is not new see e.g.,\cite{HQC} \cite{ekert}.
The point of Prop 4. is  to show explicitly how the   particular  two-qubit  Hamiltonian family associated to the untarries
(\ref{trans}) can be exploited for enacting controlled phase via a simple  adiabatic protocol.

\section{ Conclusions.}

 In this paper we analyzed the entanglement
generation capabilities of  a parametric family of adiabatically
connected non-degenerate Hamiltonians. One prepares the system in 
a separable eigenstate of of a distinguished Hamiltonian $H_0$ in the 
family and then the space of parameters is adiabatically explored. 
The system remains then in an energy eigenstate and the (bi-partite) entanglement
contained in such an eiegenstate can be maximized over the
manifold of control parameters. 
We introduced an  associated measure $e$  of adiabatic entangling
power and discussed its properties and relations with a previously introduced measure
for the case of iso-spectral families of  Hamiltonians.
We illustrated the general ideas
by studying explicitly the adiabatic entangling power of concrete
two-qubits Hamiltonian families. We also showed how to generate a
non-trivial two-qubits entangling gate by means of adiabatic
loops.

We thank M. C. Abbati, A. Mani\`a, L. Faoro and an anonymous referee for useful comments.
P.Z. gratefully acknowledges financial support by
Cambridge-MIT Institute Limited and by the European Union project  TOPQIP
(Contract IST-2001-39215)



\begin{thebibliography}{}
\bibitem{loss}Here we are assuming, without loss of generality, that Ker$ H_0=\{0\}$
\end{thebibliography}
\end{document}